\documentclass[twocolumn, twoside]{revtex4-2}
\usepackage{bm, amsmath, amsfonts, mathtools, physics}

\usepackage{graphicx}

\begin{document}

    \title{Self-Organized Bioelectricity via Collective Pump Alignment: \\
    Toward a Physical Origin of Chemiosmosis}
    
  \author{Ryosuke Nishide}
  \email{ryosuke.nishide@nbi.ku.dk}

  \author{Kunihiko Kaneko}

  \affiliation{
  Niels Bohr Institute, University of Copenhagen, Jagtvej 155 A, Copenhagen N, 2200, Denmark
  }
  
\begin{abstract}
Directional ion transport across membranes maintains living systems in nonequilibrium, which underlies chemiosmotic energy conversion. 
However, the physical origin of collectively organized ion transport in primitive cellular systems remains unclear.
Here, we propose a minimal model in which ion pumps collectively align through feedback between ion transport and electrostatic interactions.
In the model, directional ion transport generates a membrane potential, while the resulting electrochemical potential biases pump orientation, leading to self-organized collective alignment. 
Numerical simulations and mean-field analysis reveal a nonequilibrium transition from a disordered state without net transport to a pump-alignment state with sustained membrane potentials. 
The critical behavior is consistent with the mean-field Ising universality class; however, the effective field is generated self-consistently by nonequilibrium ion transport. 
We further show that protocell asymmetry can bias the polarity of the membrane potential. 
These results provide a generic self-organizing mechanism for the emergence of bioelectricity and a physical route toward chemiosmotic coupling in protocells.
\end{abstract}

\maketitle

\section*{Introduction}
Bioelectricity, arising from ion gradients across membranes, is essential for living cells\cite{schoepp2013universal}. 
Living systems actively maintain these gradients under nonequilibrium conditions.
Mitchell’s chemiosmotic theory established the coupling between ion transport and chemical reactions mediated by membranes as a central principle of bioenergetics: ion currents drive cellular energy transduction, including ATP synthesis\cite{mitchel1961coupling, mitchel1966chemiosmotic, morellil2019anupdate}. 
Through this coupling, electrochemical gradients across membranes drive intracellular chemical reactions.

The emergence of such coupling would be an essential step in the origin of life.
Phylogenetic evidence indicates that early life already synthesized ATP through membrane-potential–driven processes\cite{weiss2016physiology,moody2024nature}.
Chemical and geological studies further indicate that primitive cells may have emerged by exploiting natural ion gradients associated with alkaline hydrothermal vents to drive energetically unfavorable chemical reactions\cite{sojo2014bioenergetic, yu2025chemiosmotic}. 
However, while early systems may have relied on externally imposed ion gradients, how protocells could generate and sustain membrane potentials remains unclear.

Although ion pumps could in principle generate ion gradients, directional transport requires aligned orientation among pumps. 
If pumps are inserted with random orientations, inward and outward fluxes cancel on average, preventing the formation of a net membrane potential. 
In modern cells, membrane protein topology is regulated, but such mechanisms would not be available in early protocells. 
Moreover, fluctuations in pump orientation and electrochemical constraints associated with ion transport disrupt collective alignment, making membrane potentials difficult to establish.
This raises a fundamental question: can ion pumps collectively self-organize to establish membrane potentials in the absence of pre-existing ion gradients?
Such collective organization is essential for the emergence of chemiosmosis and bioelectricity.

From a physical perspective, this problem can be viewed as the emergence of collective pump alignment coupled to ion transport.
The origin of life has been studied from various perspectives, including autocatalytic reaction sets\cite{dyson1999origins, jain2001amodel, blokhuis2020universal, kaneko2005onrecursive}, the generation of bio-information\cite{eigen1971selforganization, kauffman1992theorigins, eigen2012thehypercycle, kaneko2002onakinetic, tkachenko2015spontaneou, falk2025suppression}, the role of lipid membranes\cite{segre2001thelipid, lancer2018systems, szathmáry2006theorigin}, and nonequilibrium conditions required for life\cite{goldenfeld2011life, branscomb2017escapement}, yet the origin of bioelectricity remains elusive. 
That is, the physical mechanism by which directional ion transport, and hence membrane potentials, spontaneously emerge remains elusive.

Here, we address this question by proposing a minimal model for the self-organized emergence of a membrane potential through interacting ion pumps.
Such a scenario is plausible in protocells, where membranes were likely more fluid and protein orientation less regulated than in modern cells.
We show that randomly oriented pumps can spontaneously align through a nonequilibrium phase transition, generating a finite membrane potential without an externally imposed ion gradient.
Mean-field analysis identifies the transition condition and its mean-field critical scaling.
Beyond mean field, we show how nonequilibrium ion transport shapes fluctuations around the transition.
Finally, we show that volume and transport asymmetries act as symmetry-breaking fields that select the polarity of the membrane potential.

\begin{figure*}[t]
    \includegraphics[keepaspectratio,scale=0.9]{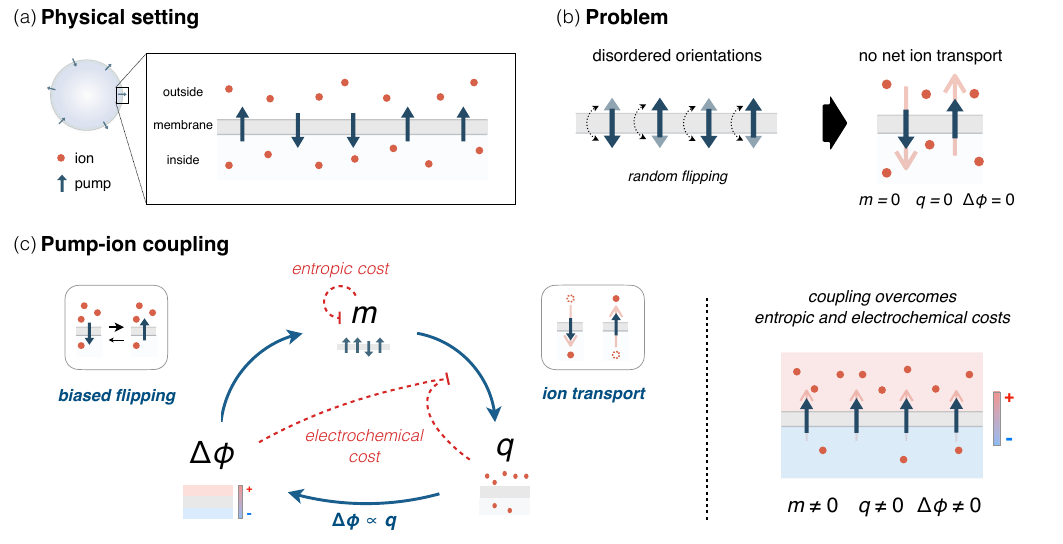}
    \caption{Minimal mechanism for self-organized membrane-potential through pump--ion coupling.
    (a) Ion pumps embedded in a membrane transport ions inward or outward depending on their orientation.
    (b) Random pump flipping favors disordered orientations, causing opposing ion fluxes cancel and the system remains without net transport; no pump alignment, $m=0$; no ion imbalance, $q=0$; and no membrane potential, $\Delta \phi=0$.
    (c) Pump alignment $m$ drives ion transport, producing an ion imbalance $q$ and hence a membrane potential $\Delta\phi$. 
    The membrane potential feeds back on pump orientation by biasing pump flipping, forming a closed feedback loop. 
    This feedback is opposed by the entropic cost of pump alignment and the electrochemical cost that suppresses further ion transport.
    When this coupling overcomes the opposing effects, pump alignment, ion imbalance, and membrane potential emerge together, yielding $m\neq0$, $q\neq0$, and $\Delta\phi\neq0$.}
\label{fig1}
\end{figure*}

\section*{Model} 
We consider a minimal model of ion pumps embedded in a membrane that collectively generate a membrane potential $\Delta\phi$ through ion transport. 
The system consists of $N_P$ pumps, each has an orientation variable $P_i=\pm1$, where $P_i=+1$ denotes outward pumping and $P_i=-1$ denotes inward pumping\footnote{Each pump is structurally asymmetric, and the direction of ion transport is determined by whether its ion-uptake side faces inward or outward\cite{morth2011astructural, dyla2020structure, calisto2021mechanisms}}.
Ions are present inside and outside the cell, with concentrations $c^I$ and $c^E$, respectively.
The total number of ions $N_Q$ is assumed to be large, allowing a continuum description of concentrations.
The pumps drive ion fluxes that determine the membrane potential, which in turn biases the pump orientations and leads to a self-consistent coupling between pump alignment and ion dynamics.

The ion flux across the membrane is driven by pumps\footnote{We neglect ion channels and leakage, assuming that ion transport is dominated by pumps; their effects do not qualitatively affect our results (see SI Appendix).} and depends on both their orientation and the electrochemical potential difference\cite{beard2005abiophysical, terradot2024escherichia, lo2024bacterial, garlid1989nature, kondepudi2014modern}. 
In particular, the flux into and out of the cell is proportional to the number of inward and outward pumps, $n_P^{\mathrm{in}}$ and $n_P^{\mathrm{out}}$.
Within an Eyring-type approximation, the flux is expressed as
\begin{align}\label{dcdt}
\partial_t c^I
&= \gamma n_P^{\mathrm{in}} c^E e^{\beta z e \phi^E}
 - \gamma n_P^{\mathrm{out}} c^I e^{\beta z e \phi^I},
\end{align}
where $\gamma$ is an effective rate constant that incorporates the energetic input required to drive the pumps, and $\phi^I$ and $\phi^E$ denote the electrostatic potentials inside and outside the cell, respectively.
For simplicity, we neglect the reverse transport against the pump orientation, which does not qualitatively affect the collective behavior.

Since $N_P$ is fixed, we define $p = n_P^{\mathrm{out}}/N_P$, so that $n_P^{\mathrm{in}} = N_P(1-p)$.
The external concentration follows from ion-number conservation as $c^E = r_V(c_0 - c^I) + c_0$, where $c_0 = N_Q/(V^I + V^E)$ is the average ion concentration and $r_V = V^I/V^E$ is the volume ratio.
The membrane potential $\Delta \phi = \phi^I - \phi^E$ is related to the ion imbalance as
\begin{align}\label{membrane_potential}
\Delta\phi = c_0 V^I \frac{z e}{C} q,
\end{align}
where $q = (c^I - c_0)/c_0$ and $C$ is the membrane capacitance.
Here we treat the membrane as a capacitor and adopt a quasi-static approximation\cite{lo2024bacterial}, so that $\Delta\phi$ is determined instantaneously by \(q\).

Using the above definitions, the ion dynamics reduce to
\begin{align}\label{dqdt}
\partial_t q
&= \gamma N_P\big\{ (1 - p)(1 - r_V q)e^{-\alpha q}
 - p(1 + q)e^{\alpha q}\big\},
\end{align}
where $\alpha = \beta c_0 V^I z^2 e^2 / (2C)$ quantifies the electrostatic energy relative to thermal energy (see Method).
The exponential factors represent the electrostatic cost associated with transport against the membrane potential.
These factors represent the electrostatic contribution to the activation barrier for transporting a charged ion from the source side of the membrane.

Next, we incorporate the interaction between the pumps and the membrane potential. 
We assume that the pump orientation is biased by the electrostatic potential across the membrane, which is plausible in early protocells with loosely structured membranes. 
Such interaction is also consistent with dynamic changes in protein topology observed in modern membranes\cite{bogdanov2018flip}.
We introduce the interaction energy as
\begin{align}
H(q,P) = zeJ \Delta\phi \sum_i P_i = \frac{2J\alpha q}{\beta} \sum_i P_i,
\label{energy}
\end{align}
where $J$ is a coupling constant, we denote the pump configuration by $P=\{P_i\}$, and in the second equality we used Eq.~(\ref{membrane_potential}). 
This establishes a self-consistent coupling between pump orientation and ion dynamics.

The time evolution of the probability distribution $\pi(P,t)$ is described by a master equation with single-pump flipping dynamics, where the transition rates depend on the instantaneous ion variable $q(t)$,
\begin{align}
    \!\!\!\partial_t\pi(P,t)
    \!&=\!\sum_l\{T(P\!\mid\! P^{l}) \pi(P^{l},t) - T(P^{l}\!\mid \!P)\pi(P,t)\},\label{dPdt}\\
    \!\!\!T(P^l\!\mid \!P)
    \!&=\!\exp\big[-\frac{\beta}{2} \{H(q,P^l)-H(q,P)\}\big],\label{transitionP}
\end{align}
where $P^l$ denotes the configuration obtained by flipping the $l$-th pump.

In the absence of coupling (e.g., $J=0$), the interaction energy vanishes ($H=0$), and the pump orientations fluctuate randomly, yielding $p=1/2$. 
Substituting this into Eq.~(\ref{dqdt}) gives $q=0$, indicating the absence of an ion gradient and hence no membrane potential.
We now examine how the coupling between pump orientation and ion dynamics leads to a feedback mechanism that can drive collective pump alignment.

Note that the transition rate in Eq.~(\ref{transitionP}) with the energy (\ref{energy}) has a structure analogous to that of the mean-field Ising model.
A key distinction, however, is that the effective field here is generated dynamically and self-consistently through nonequilibrium ion transport by the pumps themselves, as described by Eq.~(\ref{dqdt}).

The coupled dynamics of the ion concentration and the pump configuration are investigated using Eqs.~(\ref{dqdt})--(\ref{transitionP}). 
We employ a hybrid simulation approach: the pump configuration evolves stochastically via the Gillespie algorithm, while the ion concentration evolves according to an ordinary differential equation. 
The ion dynamics are therefore effectively stochastic due to their dependence on the fluctuating pump configuration. 
See SI Appendix for details of the simulation methods.

    \begin{figure}[t]
    \includegraphics[keepaspectratio,scale=0.9]{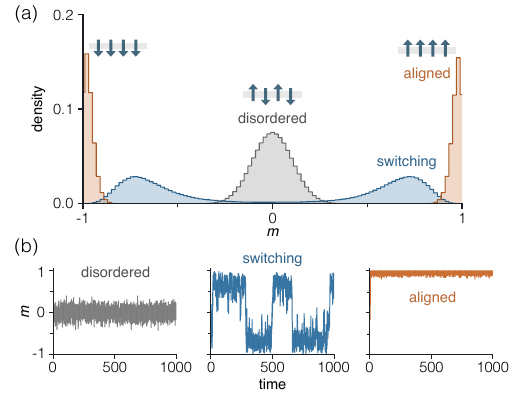}
    \caption{Three regimes. 
    Distributions of pump alignment $\langle m\rangle$ for $2000$ samples (a) and their time series for one sample (b).
    Colors correspond to states: disordered (blue,$\alpha=0.01$), switching (yellow,$\alpha=0.12$), and aligned (red,$\alpha=0.2$).
    Parameters $N_P=100,J=5.5,\gamma=1$.
    Distributions are calculated over $t=30$--$70$ and normalized so that the total amount is one.}
    \label{fig2}
    \end{figure}

\section*{Three regimes and disorder-alignment transition}
We first examine the simple case $r_V=1$ in Eq.~(3)\footnote{$q$ takes values $-1\leq q\leq1$ (i.e., from all ions outside to inside).}.
We then investigate the time evolution of pump alignment by setting $N_P=100$ and $\gamma=1$ for different values of $\alpha$. 
After a simulation time $t=10$, 
the system reaches a steady state, where three distinct regimes are observed; disordered, switching, and aligned states.
In Fig.~\ref{fig2}, histograms (for $2000$ samples) and time series (for one sample) of the pump alignment order parameter
\begin{align}\label{m}
   m=\frac{1}{N_P}\sum_iP_i=2p-1,
\end{align} 
are plotted, where $p=n_P/N_P$.

For small $\alpha$ (say $0.01$), no pump alignment is observed.
The distribution of $m$ exhibits a single peak with zero mean and fluctuations around it (Fig.~\ref{fig2}(a) and (b)).
As $\alpha$ increases (e.g., $\alpha=0.12$), the pumps are highly aligned, while the direction of alignment intermittently switches between opposite orientations (Fig.~\ref{fig2}(b)). 
Correspondingly, the distribution of $m$ is bimodal, with broad peaks around $\pm 0.7$ (Fig.~\ref{fig2}(a)).
As $\alpha$ further increases (e.g., $\alpha=0.2$), the pumps become fully aligned either inward or outward, depending on the sample.
The distribution of $m$ exhibits sharp peaks close to $\pm1$, and the alignment direction remains fixed over time for a given sample (Fig.~\ref{fig2}(b)).
These results show that increasing $\alpha$ drives the emergence of collective pump alignment.

Note that no ion gradient appears in the absence of pump alignment (SI Appendix, Fig.~S1), whereas collective pump alignment induces an ion gradient (and hence a membrane potential).
The distributions of pump alignment and ion concentration reflect the symmetry of the system under the transformation $(m,q)\rightarrow (-m,-q)$.

\begin{figure*}[t]
    \includegraphics[keepaspectratio,scale=0.9]{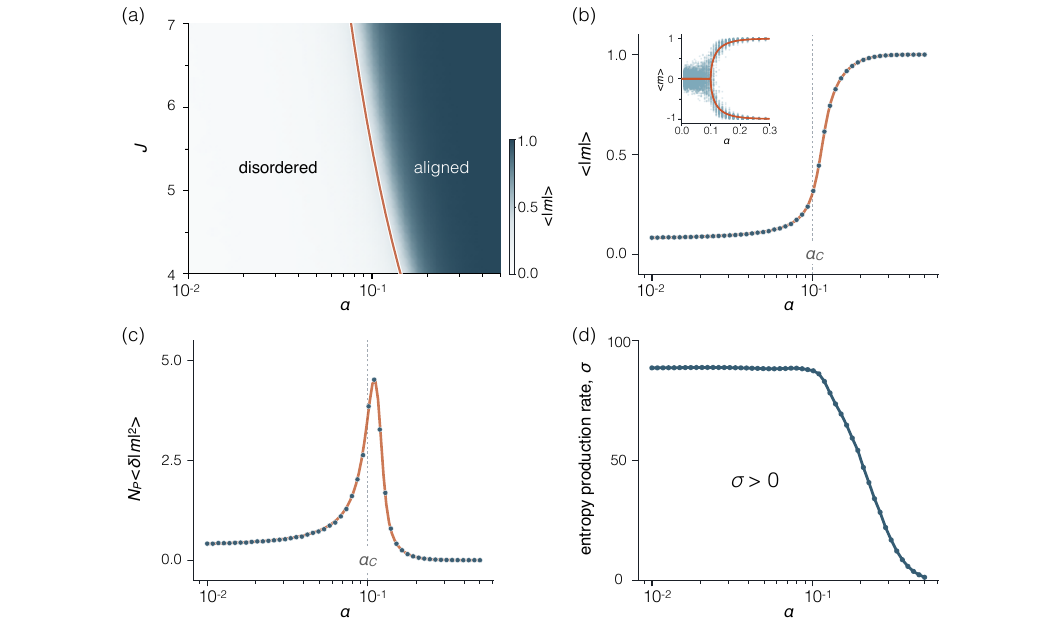}
    \caption{Transition from disorder to aligned state.
    (a) $\langle|m|\rangle$ against $J$ and $\alpha$, and transition line $J=1/(2\alpha)+1/2$ (red).
    (b) $\langle|m|\rangle$ (blue), approximation of it using Eq.~(\ref{finite distribution}) (orange), $m$ at $t=70$ (blue in inset), and stable solutions of Eq.~(\ref{selfconsistent}) (red), against $\alpha$ at $J=5.5$. 
    (c) $\langle\delta|m|^2\rangle$ multiplied by $N_P$ and approximation of it using Eq.~(\ref{finite distribution}) (orange), against $\alpha$ at $J=5.5$.
    (d) entropy production rate, against $\alpha$ at $J=5.5$.
    Parameters $N_P=100,\gamma=1$, and $200$ samples.
    The mean and variance are calculated from the samples at each time, then the time-averaged values are taken over $t=30$--$70$.}
    \label{fig3}
    \end{figure*}

To characterize the transitions among these regimes, 
we plot the ensemble mean of $|m|$\footnote{Since the pumps align bidirectionally, the state is characterized by the absolute value of $m$.}, $\langle|m|\rangle$,
and its variance $\langle\delta|m|^2\rangle$\footnote{$\delta|m|=|m|-\langle|m|\rangle$}, against $J$ and $\alpha$.
In Fig.~\ref{fig3}(a), increasing $\alpha$ or $J$ drives a transition from the disordered (white) to the aligned state (blue), with a switching regime appearing in between.
Correspondingly, the order parameter $\langle|m|\rangle$ increases sharply, while the variance $\langle\delta| m|^2\rangle$ exhibits a peak near $\alpha = 0.1$ at $J = 5.5$ (Figs.~\ref{fig3}(b) and (c) blue).
These results indicate a transition from the disordered to the aligned state, with collective pump alignment achieved beyond a phase boundary in the $(\alpha,J)$ plane.
The switching regime occurs near the critical point, characterized by a variance peak.

\section*{Mean-field analysis}
\subsection*{Self-consistent equation and transition line}
We first analyze the phase transition using the mean-field approach, motivated by the analogy between our model and the Ising model.
In the disordered and aligned states, the steady-state ion concentration exhibits peaks at zero and $\pm 1$, respectively, although $q(p)$ is affected by the fluctuation of pump orientations.
We therefore neglect fluctuations in the pump configuration and their dynamical coupling to ion transport by approximating $q(p)\simeq\bar{q}(\bar{p})$, so that the electrostatic potential acts as an effective mean field for the pumps:
\begin{align} \label{meanfield H}
   \bar{H}=\frac{2J\alpha \bar{q}}{\beta}\sum_iP_i,~\bar{q}=\bar{q}(\bar{p}),
\end{align}
where $\bar{q}$ is determined from the steady-state condition of Eq.~(\ref{dqdt}) evaluated at $p=\bar{p}$.
We then obtain a self-consistent equation for $\bar{q}$ and the pump alignment $\bar{m}=2\bar{p}-1$ (See Methods):
\begin{align} \label{selfconsistent}
   \bar{q}
   &=\tanh\{(2J-1)\alpha\bar{q}\},\\
   \bar{m}
   &=-\tanh(2J\alpha\bar{q}).
\end{align}
For $J = 0$, $(\bar{q},\bar{m}) = (0,0)$ is the only solution, showing that alignment arises from the coupling between $q$ and $m$.
The self-consistent equation exhibits a pitchfork bifurcation along the critical line
\begin{align}\label{critical line}
(2J-1)\alpha = 1.
\end{align}
For fixed $J$, this yields the critical value $\alpha_C=1/(2J-1)$.
For $\alpha<\alpha_C$, the equation has a single stable solution at $\bar{q}=0$ and $\bar{m}=0$, corresponding to the disordered state.
For $\alpha>\alpha_C$, there are three solutions: one unstable solution at $\bar{q}=0$ and two stable solutions $\bar{q}=\pm q_s$, where $q_s$ approaches one for $\alpha\gg\alpha_C$.
The two stable solutions correspond to the aligned states.
Hence, increasing $\alpha$ enhances the mutual reinforcement between the electric potential and pump alignment, inducing a transition from the fluctuation-dominated disordered phase to the aligned phase at $\alpha=\alpha_C$, with pump switching appearing as a critical behavior.
The mean-field critical line [Eq.~(\ref{critical line})] is in reasonable agreement with the numerical result (Fig.~\ref{fig3}(a)), although the transition is slightly shifted beyond $\alpha_C$.

From Eq.~(\ref{critical line}), the transition condition $2J\alpha>\alpha+1$ can be understood in terms of competing biases in the transition: the alignment bias $(2J\alpha)$ favors collective pump alignment, whereas $\alpha$ (from the membrane potential) and $1$ (from entropy) favor disorder.
This reflects the fact that macroscopic pump alignment requires the energy gain from pump flipping to exceed the electrochemical potential costs associated with ion transport (see Methods).
Furthermore, Eq.~(\ref{critical line}) shows the absence of a phase transition for $J<1/2$ (SI Appendix, Fig.~S2), where the alignment bias can never overcome the transport-induced cost.

The self-consistent equation for $\bar q$ takes the same form as that of the mean-field Ising model. 
However, the pump alignment $\bar{m}$ is determined through an additional relation involving both $J$ and $\alpha$. 
Thus, the bifurcation structure is Ising-like, but the ordering field is not an equilibrium interaction field; it is generated self-consistently by ion transport and constrained by the membrane potential. 
In the limit of large $J$ and small $\alpha$ with fixed $J\alpha$, the model reduces to the standard mean-field Ising case.

\subsection*{Critical scaling and universality}
Near $\alpha = \alpha_C$, expanding Eq.~(\ref{selfconsistent}) around $\bar{q}=0$ yields the following critical behavior (see Methods):
\begin{align}
  \bar{q}&\sim \pm \sqrt{3}\Delta \alpha^{\frac{1}{2}},\label{critical q}\\
  \bar{m}&\sim\mp\sqrt{3}(1+\alpha_C)\Delta \alpha^{\frac{1}{2}}\label{critical m},
\end{align}
where $\Delta \alpha=(\alpha-\alpha_C)/\alpha_C>0$. 
These results yield the mean-field critical exponent $1/2$, arising from the $\bar{q} \to -\bar{q}$ symmetry of the self-consistent equation, placing the system in the mean-field Ising universality class.
The self-consistent solution is plotted as the orange line in inset of Fig.~\ref{fig3}(b), showing that the numerical results follow the mean-field critical scaling.
Furthermore, as shown in SI Appendix, Fig.~S3, the slope of $\langle \bar{m} \rangle$ near the transition becomes steeper as $N_P$ increases, and the transition point approaches $\alpha_C$. 
This indicates that the deviation from the mean-field analysis arises from finite-size effects and vanishes in the large-$N_P$ limit.

Despite its nonequilibrium nature, the critical behavior falls into the mean-field Ising universality class. 
This indicates that the universal scaling is governed primarily by the symmetry and structure of the self-consistent equation rather than the microscopic origin of the effective field.

\section*{Nonequilibrium effects}
The mean-field analysis does not capture fluctuation effects such as the variance peak near the transition, because fluctuations in pump alignment dynamically feed back through nonequilibrium ion transport. 
In addition, the finite-size shift of the variance peak suggests that ion transport dynamics also modify the effective transition bias.

This nonequilibrium nature is reflected in a positive entropy production rate (EPR), a measure of irreversibility (Fig.~\ref{fig3}(d)).
The EPR is higher in the disordered regime, where frequent pump flipping perturbs the ion imbalance and dissipates the input energy through incoherent transport.
In contrast, in the aligned regime, the membrane potential is maintained by coherent pump alignment, and ion transport primarily serves to sustain this state, leading to a lower, but still positive, EPR.
Thus, the aligned state is not an equilibrium ordered state, but a nonequilibrium steady state sustained by active transport.

To incorporate dynamical effects beyond the mean-field analysis, we consider an approximate description in which ion transport relaxes rapidly compared with pump alignment dynamics. 
This leads to $\partial_t q\simeq0$, so that the ion concentration is constrained by the instantaneous pump configuration through a relation $q^*(p)$. 
The resulting effective pump--ion interaction Eq.~(\ref{energy}) is then given by
\begin{align}\label{fast ion H}
   H^* = \frac{2J\alpha q^*}{\beta}\sum_i P_i, 
   \quad \partial_t q|_{q=q^*(p)} = 0.
\end{align}

This reduction yields an effective one-variable dynamics for pump alignment. 
In the absence of cyclic transitions, the reduced dynamics satisfy detailed balance, allowing the stationary distribution to be constructed analytically. 
Although this approximation loses the full irreversibility of the original dynamics, it retains the pump-configuration dependence of the ion response through \(q^*(p)\).
The ratio of forward and backward transitions for pump flipping is given by
\begin{align}\label{R}
R(p)=\frac{p+\Delta p}{1-p}\exp\left[2J\alpha\{q^*(p)+q^*(p+\Delta p)\}\right],
\end{align}
which determines the effective bias for the transition $p \to p+\Delta p$, where $\Delta p=1/N_P$.
The first factor represents the finite-size entropic contribution, whereas the exponential factor represents the bias induced by ion transport.
Because detailed balance holds in the reduced one-variable dynamics, the stationary distribution satisfies $\pi^*(p)/\pi^*(p+\Delta p)=R(p)$, 
\begin{align}\label{finite distribution}
    \pi^*(p)\propto \exp\{-\sum_{p}^{p-\Delta p} \ln R(p')\},
\end{align}
from which the mean and variance of pump alignment can be evaluated.
This reduced description reproduces the finite-size shift of the variance peak observed in numerical simulations (Fig.~\ref{fig3}(b) and (c)), which is not captured by the mean-field theory.

Importantly, the dependence on $q^*(p)+q^*(p+\Delta p)$ in Eq.~(\ref{R}) originates from the finite change in the ion response caused by a single pump-flipping step. 
This term is absent in the mean-field description, where the ion response is evaluated only at the average pump alignment. 
Because pump flipping modifies the stationary ion response itself, finite systems retain a correction from the pump-configuration-dependent ion response. 
This correction modifies the effective stationary distribution of pump alignment and can shift finite-size observables, such as the variance peak, away from the mean-field critical point.

This correction vanishes in the large-$N_P$ limit. 
The corresponding fluctuation scaling approaches the mean-field Ising form (see Methods).

\section*{Alignment polarity under symmetry-breaking field}
\subsection*{Volume difference $r_V$}
We investigate the effect of the volume ratio $r_V$ using Eq.~(\ref{dqdt}), including a backward reaction term\footnote{For small $r_V$, $q$ can take large values ($-1\leq q\leq 1/r_V$), so that the backward reaction becomes non-negligible. Here we use a backward pumping rate, $\gamma^b=10^{-5}$.}. 
In this case, the system still undergoes a phase transition from the random state to the aligned state as $\alpha$ increases, but the aligned state becomes asymmetric with respect to the pump direction. 

\begin{figure}[t]
    \includegraphics[keepaspectratio,scale=0.9]{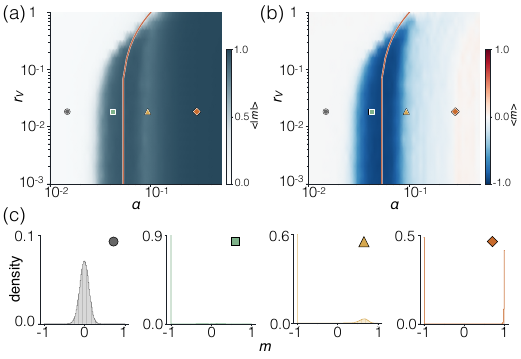}
    \caption{$r_V$ dependence of pump alignment.
    (a) $\langle m\rangle$ against $r_V$ and $\alpha$.
    (b) $\langle |m|\rangle$ against $r_V$ and $\alpha$.
    The orange lines indicate the bifurcation line $r_V=2\alpha(2J/\eta-1)-1$.
    (c) Distributions at four points.
    Parameters $N_P=100,\gamma=1,J=5.5$, and $200$ samples.
    The mean is calculated from the samples at each time, then the time-averaged values are taken over $t = 0$--$100$.}
    \label{fig4}
    \end{figure}

To characterize this symmetry breaking, we plot both $\langle |m| \rangle$ and $\langle m \rangle$ in the $\alpha$--$r_V$ plane (Fig.~\ref{fig4}(a) and (b)). 
We further show the corresponding distributions for four representative parameter sets in Fig.~\ref{fig4}(c).
For a fixed $r_V$, the system is initially in the disordered state for small $\alpha$ (Fig.~\ref{fig4}(a)--(c), circle). 
Increasing $\alpha$ first produces inward pump alignment (Fig.~\ref{fig4}(a)--(c), square). 
With further increase in $\alpha$, aligned states in both directions emerge (Fig.~\ref{fig4}(a)--(c), triangle and diamond).

To understand this behavior, we analyze the steady states based on the self-consistent equation using a mean-field approximation (see Methods). 
This leads to
\begin{align}\label{SC r_V}
   \bar{q}
   &\simeq\tanh\big\{(2J-1)\alpha\bar{q}+\frac{1}{2}\ln\frac{1-r_V \bar{q}}{1-\bar{q}}\big\},
\end{align}
where the logarithmic term explicitly breaks the symmetry under $\bar{q}\to -\bar{q}$, for $r_V\neq1$
For $r_V<1$, the exterior contains a larger number of ions than the interior. 
As a result, inward pumping generates a larger membrane potential than outward pumping, thereby favoring inward alignment.
Only when $\alpha$, and hence the electrostatic contribution, becomes sufficiently large do aligned states in both directions become comparably stabilized.
This balance corresponds to the transition at
\begin{align}\label{critical alpha rV}
\alpha_C=\frac{1+r_V}{2}\frac{\eta}{2J-\eta}.
\end{align}

In numerical simulations, inward alignment remains preferentially observed slightly beyond $\alpha_C$, since the outward aligned solution remains close to $\bar q=0$ near the transition and is strongly affected by fluctuations.

Furthermore, decreasing $r_V$ extends the asymmetric region toward smaller $\alpha$ and suppresses switching between aligned directions by enhancing the asymmetry of the membrane-potential response.

\subsection*{Transport-rate difference $r_\gamma$}
So far, we have considered a symmetric situation without an explicit directional bias between the inside and outside, and examined whether directional pump alignment can emerge spontaneously through ion transport. 
Now we consider the case where the inward and outward transport rates, $\gamma^{\mathrm{in}}$ and $\gamma^{\mathrm{out}}$, differ, which may originate from asymmetric chemical conditions between the interior and exterior.
In this case, the corresponding self-consistent equation is given by
\begin{align}\label{SC r_gamma}
    \bar{q}
    =\tanh\big\{(2J-1)\alpha \bar{q}+\frac{1}{2}\ln r_\gamma\big\},
\end{align}
which no longer passes through the origin, resulting in asymmetric alignment and hysteresis (SI Appendix, Fig.~S4). 
Thus, $r_\gamma$ acts analogously to an explicit symmetry-breaking field in the Ising model.

Unlike the asymmetry induced by the volume ratio $r_V$, which preserves the $\bar q=0$ solution and induces concentration-dependent deviation, $r_\gamma\neq1$ explicitly shifts the steady state away from the origin through a constant bias term proportional to $(\ln r_\gamma)/2$ (see Method).

\section*{Discussion}
We demonstrated that membrane potentials and directional ion transport can emerge spontaneously through collective pump alignment under nonequilibrium conditions. 
This collective organization provides a generic mechanism for the emergence of chemiosmosis and bioelectricity in protocells.
As the pump--ion coupling ($J$) or electrostatic coupling ($\alpha$) increases, the system undergoes a transition from a disordered state without net ion transport to a collectively aligned state with a finite membrane potential.
The transition is well captured by a mean-field analysis analogous to the Ising model, despite the intrinsically nonequilibrium nature of the underlying ion transport dynamics.

However, the mean-field description does not capture fluctuation effects such as the variance peak and the finite-size shift of the transition.
These effects arise from the dynamical response of ion transport. 
In particular, pump flipping modifies the ion response itself, generating corrections to the effective transition bias even within the quasi-equilibrium description.
As a result, the transition is shifted from the mean-field prediction and fluctuations are enhanced near the transition point, showing that signatures of the underlying nonequilibrium transport dynamics remain encoded in the effective static transition structure.

We further showed that asymmetry in volume or transport rates biases collective pump alignment, leading to directional membrane potentials. 
In particular, volume asymmetry favors inward alignment over an intermediate parameter range, suggesting that finite protocell geometry may have contributed to the selection of transport polarity in early systems. 
Transport asymmetry similarly acts as a symmetry-breaking field, stabilizing one alignment direction over the other.

These results provide a possible physical scenario for how membrane potentials emerged in protocells.
While externally-supplied ion gradients have been proposed as a driving force for early bioenergetics\cite{sojo2014bioenergetic}, our results suggest that protocells could instead spontaneously generate membrane potentials and directional ion transport through collective ion--pump interactions, even in the absence of pre-existing gradient.
Externally supplied gradients could nevertheless act as additional symmetry-breaking fields, together with the volume and transport asymmetries demonstrated here.
Moreover, since collective alignment is initiated by fluctuations, stochastic dynamics associated with a finite number of pumps may have played an important role in early protocells, where the number of pumps would be limited. 
How these fluctuation-driven alignment dynamics depend on ion transport timescales and pumping kinetics, including the role of $\gamma$, remains an important direction for future study.
Furthermore, changes in membrane composition or pump properties could alter the effective couplings $J$ and $\alpha$, allowing protocells to cross a critical point, and leading to the emergence of bioelectricity.

The coupling between pump orientation and membrane potential assumed in our model may be particularly relevant for primitive membranes, where protein topology would not have been yet tightly controlled. 
Also, even in modern membranes, experiments have shown that membrane protein topology can dynamically reorganize in response to charge redistribution\cite{bogdanov2018flip}. 
These observations raise possibility that electrostatic effects contributed to the emergence of collective pump alignment in primitive membranes and may also be related to the positive-inside rule\cite{von2006membrane} observed in modern cells.

Here we presented a minimal model for the self-organized emergence of membrane potential through collective ion transport. 
Extending the model to multiple ion species and explicit pump kinetics\cite{calisto2021mechanisms, beard2005abiophysical, terradot2024escherichia} will further clarify the emergence of membrane potentials in protocells. 
More generally, the pumping rate itself can depend on nonequilibrium chemical reactions coupled to membrane potentials, generating additional feedback between transport, reaction dynamics, and bioelectric organization. 
Also, incorporating ion channel, leakage, and saturation together with interfacial electrostatics beyond the present capacitor approximation may further clarify the physical principles governing protocell bioelectricity.

Our results demonstrates that collective ion--pump interaction provides a generic nonequilibrium mechanism for the spontaneous emergence of membrane potentials, bioelectricity, and early bioenergetics. 
More broadly, our work highlights how collective organization driven by nonequilibrium transport can generate functional biological states even without externally imposed asymmetry. 
Our results will shed light on how chemiosmotic coupling and bioelectric organization emerged in protocells.

\section*{Acknowledgement}
The authors would like to thank Riz Noronha and Shunsuke Ichii for useful discussions.
This study was supported by the Novo Nordisk Foundation Grant No.~NNF21OC0065542.

\section*{Methods}
\subsection*{Membrane potential}
The membrane potential $\Delta\phi$ arises from an imbalance of charges across the membrane, which has low ion permeability and acts as a capacitor\cite{lo2024bacterial}.
It is defined as the difference between the net charges inside, $\mathcal{Q}^I_{\mathrm{net}}$, and outside, $\mathcal{Q}^E_{\mathrm{net}}$, the cell:
\begin{align}
    \Delta\phi 
    &=\frac{1}{C}(\mathcal{Q}^I_{\mathrm{net}}-\mathcal{Q}^E_{\mathrm{net}})
    =\frac{e}{C}\sum_i(z_in_{i}^I-z_in_{i}^E),
\end{align}
where $n_i^{I}$ and $n_i^{E}$ is the numbers of internal and external ions of type $i$, respectively, $z_i$ is the valence of ions $i$, and $C$ is the membrane capacitance.
Assuming only ions $Q$ can transport across the membrane and other ions have no gradient,
\begin{align}
   c_j\equiv\frac{n_j^I}{V^I}=\frac{n_j^E}{V^E}, \quad {}^\forall j~ \neq Q,
\end{align}
the net charges satisfy $\mathcal{Q}^I_{\mathrm{net}} = -\mathcal{Q}^E_{\mathrm{net}}$. 
With the total number $N_Q = n_Q^I + n_Q^E$, the membrane potential can be expressed as
\begin{align}
\Delta\phi = \frac{z_Q e}{C}\left(n_Q^I - \frac{V^IN_Q}{V^I+V^E}\right)
= \frac{z_Q e V^I}{C}(c^I - c_0),
\end{align}
where $c^I = n_Q^I / V^I$ is the internal concentration of $Q$ and 
$c_0 = N_Q / (V^I+V^E)$ is the average concentration.
Internal and external potentials relative to the membrane surface are then \begin{align}
    \phi^I=-\phi^E=\frac{\Delta\phi}{2}.
\end{align}

\subsection*{Mean-field analysis}
We derive the self-consistent equation~(\ref{selfconsistent}) using a mean-field analysis.
In the steady state, we define $\pi^*(p)$ as the steady distribution of $p$,
and $q^*(p)$ as the steady solution of Eq.~(\ref{dqdt}).
For large $N_P$, $\pi^*(p)$ is sharply concentrated around its stable peak(s).
Above the transition, the stationary distribution becomes bimodal, with two peaks related by
$(p,q)\to(1-p,-q)$.
Thus, around each peak, we approximate
$q^*(p)\simeq q^*(\bar p)\equiv \bar{q}$,
where $\bar{p}$ is the peak position of the selected branch, neglecting fluctuations within the peak to leading order.

The quantities $\bar p$ and $\bar q$ therefore denote representative values within the selected branch.
For $r_V=1$, the steady-state condition of Eq.~(\ref{dqdt}) then gives
\begin{align} \label{meanfield dqdt=0}
    \bar{p}
    =\frac{1-\bar{q}}{(1+\bar{q})e^{2\alpha \bar{q}}+1-\bar{q}} .
\end{align}

Using $\bar q$ as the mean field, the energy~(\ref{energy}) reduces to the form given in Eq.~(\ref{meanfield H}). 
For a fixed $\bar q$, the corresponding distribution is given by 
\begin{align}
    \bar{\pi}(P)=\frac{e^{-\beta \bar{H}(P)}}{Z},
    \quad 
    Z=\{2\cosh(2J\alpha \bar{q})\}^{N_P}.
\end{align}
The mean-field pump orientation is then
\begin{align} \label{m mean}
    \bar m
    =2\bar{p}-1=
    -\frac{1}{2J\alpha N_P}
    \frac{\partial \ln Z}{\partial \bar{q}}
    =-\tanh(2J\alpha\bar{q}).
\end{align}

Using Eqs.~(\ref{meanfield dqdt=0}) and (\ref{m mean}),
we obtain the self-consistent equation for $\bar{q}$:
\begin{align} 
   \tanh(2J\alpha\bar{q})
   =\tanh(\alpha\bar{q}+\tanh^{-1}\bar{q}).
\end{align}
Since $\tanh(\cdot)$ is a monotonic function, this yields Eq.~(\ref{selfconsistent}).
Thus, the bifurcation line is given by Eq.~(\ref{critical line}).
Therefore, for $(2J-1)\alpha<1$, there is one stable solution, $\bar{q}=0$, whereas for $(2J-1)\alpha>1$, there are two stable solutions, $\bar{q}=\pm q_s$, and one unstable solution, $\bar{q}=0$.

\subsection*{Interpretation of the transition condition}
From Eq.~(\ref{critical line}), the alignment phase does not appear for $J<1/2$.
This can be understood in terms of the competition between ion transport and pump switching in the random phase.
We describe the ion transport using the effective chemical 
potentials $\mu^{I/E}$ and the effective work terms $w^{I/E}$, as
\begin{align}
    \partial_t q=e^{\mu^E+w^E}-e^{\mu^I+w^I},
\end{align}
where, for simplicity, we set $\beta=1$ and $\bar{\gamma}=\gamma N_P$, and
\begin{align}
    \begin{cases}
    \mu^E=\ln(1-q)-\alpha q,
    &w^E=\ln\{\bar{\gamma}(1-p)\},\\
    \mu^I=\ln(1+q)+\alpha q,
    &w^I=\ln(\bar{\gamma} p).
    \end{cases}
\end{align}
The total chemical potential associated with the ion transport into the cell is
\begin{align}
    \mu_{\mathrm{ion}}
    &=\mu^I+w^I-(\mu^E+w^E)\nonumber\\
    &=\ln\frac{1+q}{1-q}+2\alpha q+\ln\frac{p}{1-p}.
\end{align}
For pump switching, the chemical potential is
\begin{align}
    \mu_{\mathrm{flip}}
    &=\ln\frac{1-p}{p}+4J\alpha q,
\end{align}
where a positive $\mu_{\rm flip}$ favors a flip that decreases $p$, and vice versa. 
Note that we here consider the large $N_P$ limit, treat $p$ as a continuous variable.
At the phase transition, the solution $(q,p)=(0,1/2)$ needs to be unstable, and is replaced by the aligned state.
To see the condition for this to occur, consider a microstate in which the ion concentration has increased by an infinitesimal amount $dq>0$, i.e., $(q,p)=(dq,1/2)$. 
At this state, there are two competing processes: (i) an inward pump flip, and (ii) a return of the ions, which would bring $q$ back to zero. 
The driving force associated with an inward pump flip is $d\mu^\star_{\rm flip} = 4 J \alpha dq$.
In contrast, the chemical potential decrease associated with returning the ions is $d\mu_{\rm ion}^\star = (2 + 2\alpha) dq$.
Comparing the magnitudes of these chemical potentials, the macroscopic alignment phase occurs only if the pump flip is favored over ion return, i.e.,
\begin{align}
d\mu_{\rm flip}^\star > d\mu_{\rm ion}^\star \quad \Rightarrow \quad J > \frac{1}{2} + \frac{1}{2\alpha}.
\end{align}
%As $\alpha > 0$, the lower bound reduces to $1/2$.
As $\alpha>0$, the right-hand side is always larger than $1/2$ and approaches $1/2$ only in the limit $\alpha\to\infty$. 
Thus, no transition is possible for $J<1/2$.

\subsection*{Critical exponent}
In the vicinity of $\bar{q}=0$, we expand the left-hand side and right-hand side of self-consistent equation~(\ref{selfconsistent}) as
\begin{align}
  \mathrm{LHS}
    &=\frac{1+\alpha_C}{\alpha_C}\alpha\bar{q}-\frac{1}{3}\big(\frac{1+\alpha_C}{\alpha_C}\big)^3\alpha^3\bar{q}^3+\mathcal{O}(\bar{q}^5),\\
 \mathrm{RHS}
    &=(1+\alpha)\bar{q}
    -\big(\alpha+\alpha^2+\frac{1}{3}\alpha^3\big)\bar{q}^3+
   \mathcal{O}(\bar{q}^5),
\end{align}
where we used $J=1/(2\alpha_C)+1/2$.
Up to third order, the nonzero solutions are given by
\begin{align}
 \!\!\bar{q}^2
  \simeq\frac{\alpha-\alpha_C}{\alpha_C}\big\{\big(\frac{1}{3\alpha_C^3}
  +\frac{1}{\alpha_C^2}
  +\frac{1}{\alpha_C}\big)\alpha^3
  -\alpha^2-\alpha\big\}^{-1}.
\end{align}
Thus, for small $\Delta\alpha=(\alpha-\alpha_C)/\alpha_C>0$, retaining only the leading-order contribution yields Eqs.~(\ref{critical q}) and (\ref{critical m}).

\subsection*{Entropy production rate}
We quantify the nonequilibrium character of the steady state by the entropy
production rate associated with pump-flipping transitions. 
Let $\pi(m,q)$ denote the stationary joint distribution of $m$ and $q$, and let $\Delta m=2/N_P$. 
For fixed $q$, we denote by $W_+(m,q)$ the rate for the transition $m\to m+\Delta m$, and by $W_-(m,q)$ the rate for the transition $m\to m-\Delta m$.
Explicitly,
\begin{align}
    W_+(m,q)
    &= N_P\frac{1-m}{2}\exp(-2J\alpha q),\\
    W_-(m,q)
    &=N_P\frac{1+m}{2}\exp(2J\alpha q).
\end{align}
The probability current between neighboring values of $m$ is then
\begin{align}
  J_{m}(q) =&\quad W_+(m,q)\pi(m,q)\nonumber\\
   &-W_-(m+\Delta m,q)\pi(m+\Delta m,q).
\end{align}
The entropy production rate is computed as
\begin{align}
    \sigma
    =\sum_m \int dq\,
    J_{m}(q)
    \ln\frac{W_+(m,q)\pi(m,q)}{W_-(m+\Delta m,q)\pi(m+\Delta m,q)} .
\end{align}
In numerical simulations, the stationary distribution $\pi(m,q)$ was estimated from time-weighted histograms over the steady-state time window. The continuous variable $q$ was discretized into bins, and the integral over $q$ was evaluated as a sum over these bins.

\subsection*{Fast ion transport approximation}
In the fast ion transport approximation, the ion imbalance is assumed to relax
rapidly compared with pump flipping. 
Thus, for a given pump configuration, the ion variable is constrained by the steady-state condition $\partial_t q=0$.
For $r_V=1$, Eq.~(\ref{dqdt}) gives
\begin{align} \label{fast q p relation}
    p=\frac{1-q^*}{(1+q^*)e^{2\alpha q^*}+1-q^*} .
\end{align}
Substituting $q=q^*(p)$ into the energy~(\ref{energy}) gives the effective
energy in Eq.~(\ref{fast ion H}).

Eliminating the fast variable $q$ in this way yields an effective one-dimensional dynamics for $p$, with nearest-neighbor steps $\Delta p=1/N_P$.
Along the fast ion branch $q=q^*(p)$, the corresponding transition rates are
\begin{align}
    W_+(p)&=N_P(1-p)\exp[-2J\alpha q^*(p)],\\
    W_-(p+\Delta p)&=N_P(p+\Delta p)\exp[2J\alpha q^*(p+\Delta p)],
\end{align}
where $W_+(p)$ denotes the rate for $p\to p+\Delta p$, and
$W_-(p+\Delta p)$ denotes the reverse rate $p+\Delta p\to p$.
Their ratio $R(p)\equiv W_-(p+\Delta p)/W_+(p)$ corresponds to Eq.~(\ref{R}).

Because the reduced process is one-dimensional with reflecting boundaries, there are no cyclic currents. 
The steady-state current therefore vanishes, and detailed balance holds, $R(p)=\pi^*(p)/\pi^*(p+\Delta p)$.
For finite $N_P$, the stationary distribution is obtained recursively as Eq.~(\ref{finite distribution}).
The mean and variance of pump alignment are evaluated from this distribution.

In the large-$N_P$ limit, $q^*(p+\Delta p)\simeq q^*(p)$, and the distribution
takes the large-deviation form
\begin{align}
    \pi^*(p)
    &\propto e^{-N_P\Phi(p)},\\
    \Phi(p)
    &=\int^p dp'\{\ln\frac{p'}{1-p'}+4J\alpha q^*(p')\}.
\end{align}
Expanding $\Phi(p)$ around its stable minimum gives
\begin{align} \label{fast p var}
    \langle \delta p^2\rangle
    =
    \frac{1}{N_P}
    \frac{1}
    {
    \frac{1}{p(1-p)}
    +
    4J\alpha\frac{dq^*}{dp}
    }
    =
    \frac{1}{4}\langle \delta m^2\rangle .
\end{align}
From Eq.~(\ref{fast q p relation}), we obtain
\begin{align} \label{fast m var}
&N_P\langle \delta m^2\rangle\nonumber\\ 
&=\frac{4e^{2\alpha q^*}\{1-q^{*2}\}
\{1+\alpha(1-q^{*2})\}}
{\{(1+q^*)e^{2\alpha q^*}+1-q^*\}^2
\{1-\alpha(2J-1)(1-q^{*2})\}}.
\end{align}
On the disordered side near the critical point, where $q^*=0$ and $\alpha_C(2J-1)=1$, we obtain
\begin{align} \label{fast m var critical}
N_P\langle \delta m^2\rangle
\simeq
(1+\alpha_C)|\Delta\alpha|^{-1}.
\end{align}
Thus, the variance follows the mean-field critical exponent $-1$.

\subsection*{Result for $r_V\neq1$}
The ion concentration dynamics obtained by adding the backward reaction to Eq.~(\ref{dqdt}) is given by
\begin{align}
   \partial_tq
    =&~\quad\gamma N_P \{(1-p)(1-r_Vq)e^{-\alpha q}
    - p(1+q)e^{\alpha q}\}\nonumber\\
    &-\gamma^b N_P \{(1-p)(1+q)e^{\alpha q}
    + p(1-r_Vq)e^{-\alpha q}\}.
\end{align}
Using the mean-field approximation, we obtain the self-consistent equation:
\begin{align}
   \tanh(2J\alpha\bar{q})
   &=\eta\tanh(\alpha\bar{q}+\frac{1}{2}\ln\frac{1-r_V \bar{q}}{1-\bar{q}}).
\end{align}
Since $\eta=(\gamma+\gamma^b)/(\gamma-\gamma^b)
=1+\mathcal{O}\{(\gamma^b/\gamma)^2\}\simeq1$
and $\tanh(\cdot)$ is monotonic, we obtain Eq.~(\ref{SC r_V}).
This is consistent with the self-consistent equation of the mean-field Ising model with an effective external field that depends self-consistently on $\bar q$.

The logarithmic term explicitly breaks the symmetry under $\bar q\to-\bar q$, leading to asymmetric bifurcation. 
Fixing $r_V$ and increasing $\alpha$, the system develops stable fixed points at $\bar q=0$ and $\bar q=q_+>0$, separated by an unstable fixed point at $\bar q=q_-<0$. 
With further increase in $\alpha$, the $\bar q=0$ state loses stability and $\bar q=q_-$ becomes stable through a transcritical-like bifurcation.
This transition line is determined from the stability condition of the $\bar q=0$ solution. 
Evaluating the derivative of the left-hand side minus the right-hand side of Eq.~(\ref{SC r_V}) at $\bar q=0$ yields Eq.~(\ref{critical alpha rV}).
For $\alpha>\alpha_C$, both inward ($\bar q=q_+$) and outward ($\bar q=q_-$) aligned states are stable.

\subsection*{Result for $r_\gamma\neq1$}
When the inward and outward
transport rates differ, 
the ion concentration dynamics is given by
\begin{align}
    \!\!\!\!\partial_tq
     \!= \!\!N_P\{ \gamma^{\mathrm{in}}(1-p)(1-q)e^{-\alpha q}
    -\gamma^{\mathrm{out}} p(1+q)e^{\alpha q}\}.
\end{align}
Using the mean-field approximation, obtain the self-consistent equation:
\begin{align} 
 \!\!\!\! \tanh(2J\alpha \bar{q})
  =\tanh(\tanh^{-1}\bar{q}+\alpha \bar{q}-\frac{1}{2}\ln r_\gamma).
\end{align}
Since $\tanh(\cdot)$ is a monotonic function, which yields
Eq.~(\ref{SC r_gamma}).
This consistent with the self-consistent equation of mean-field Ising model under the external field $(\ln r_{\gamma})/2$.

%\clearpage

% Bibliography
\bibliography{bib}

\end{document}